\documentclass[12pt]{article}
\usepackage[letterpaper,top=2cm,bottom=2cm,left=3cm,right=3cm]{geometry}
\usepackage{graphicx}
\usepackage[numbers,sort&compress]{natbib}
\usepackage{float}
\usepackage{amsmath,amssymb,amsfonts}
\usepackage{amsthm}
\usepackage{booktabs,multirow}
\usepackage[dvipsnames]{xcolor}
\usepackage[colorlinks=true,allcolors=blue]{hyperref}
\usepackage{cleveref}
\usepackage{array}
\usepackage{authblk}
\usepackage[
        range-phrase=--,
        range-units=single,
        retain-explicit-plus=true,
        retain-unity-mantissa=false,
    ]{siunitx}

\newcommand{\SHG}{\SI{1065.35}{}}
\newcommand{\SHGerr}{\SI{51.14}{\%/(W \ cm^{2})}}

\newcommand{\maxsqz}{\SI{-0.66}{dB}}
\newcommand{\maxasqz}{\SI{+13.22}{dB}}
\newcommand{\maxsqzerr}{\SI{0.35}{dB}}
\newcommand{\maxasqzerr}{\SI{0.44}{dB}}

\newcommand{\pump}{\SI{205}{mW}}
\newcommand{\maxsqzonchip}{\SI{-18.06}{}}
\newcommand{\maxasqzonchip}{\SI{+20.64}{}}
\newcommand{\maxsqzonchiperr}{\SI{0.43}{dB}}
\newcommand{\maxasqzonchiperr}{\SI{0.35}{dB}}

\newcommand{\visible}{\SI{786.1}{nm}}
\newcommand{\nir}{\SI{1572.2}{nm}}

\newcommand{\proploss}{0.1}
\newcommand{\hide}[1]{}

\theoremstyle{plain}

\theoremstyle{definition}

\theoremstyle{remark}

\newcommand*\samethanks[1][\value{footnote}]{\footnotemark[#1]}

\title{18-dB on-chip vacuum squeezing in an adaptively poled lithium niobate waveguide}

\author[1,4]{Tushar Sanjay Karnik\thanks{These authors contributed equally.}}
\author[1,2]{Xinyi Ren\samethanks}
\author[1,2]{Chun-Ho Lee\samethanks}
\author[3,6]{Bo-Han Wu}
\author[1]{Mihir Chaudhari}
\author[1,2]{Clayton Cheung}
\author[1]{James Wang}
\author[3]{Shi-Yuan Ma}
\author[3]{Mahmoud Jalali Mehrabad}
\author[1]{Ian Christen}
\author[1,2]{Reshma Kopparapu}
\author[1,4]{Kiwon Kwon}
\author[1,2,4]{Yue Yu}
\author[3]{Sri Krishna Vadlamani}
\author[1,2]{Kamila Kunes}
\author[2,5]{Quntao Zhuang}
\author[3]{Dirk Englund}
\author[1,2]{Zaijun Chen}
\author[1,2,4]{Mengjie Yu\thanks{Corresponding author: \href{mailto:mengjie.yu@berkeley.edu}{mengjie.yu@berkeley.edu}}}

\affil[1]{Department of Electrical Engineering and Computer Sciences, University of California, Berkeley, CA 94720, USA}
\affil[2]{Ming Hsieh Department of Electrical and Computer Engineering, University of Southern California, Los Angeles, CA 90089, USA}
\affil[3]{Research Laboratory of Electronics, MIT, Cambridge, MA 02139, USA}
\affil[4]{Materials Sciences Division, Lawrence Berkeley National Laboratory, Berkeley, CA 94720, USA}
\affil[5]{Department of Physics and Astronomy, University of Southern California, Los Angeles, CA 90089, USA}
\affil[6]{Department of Electrical and Computer Engineering, University of Hawaii at M\={a}noa, Honolulu, HI 96822, USA}

\date{} % no date on arXiv

\begin{document}
% \linenumbers
\maketitle

\begin{abstract}
Quantum squeezed states of light can enhance measurement sensitivity beyond classical limits and enable quantum information processing, but scalable low-loss sources remain challenging. We demonstrate continuous-wave quantum squeezing on a chip, achieving 18 dB of squeezing and 20 dB of anti-squeezing at 1570 nm in a 1.6-cm traveling-wave adaptively poled thin-film lithium niobate waveguide. A distributed model independently determines facet losses, phase noise, and nonlinear interaction strength without prior assumptions, enabling rigorous inference of on-chip performance. We estimate a 95$\%$ confidence interval of [-18.96, -17.25] dB squeezing and [19.96, 21.35] dB anti-squeezing. These values represent the highest squeezing reported for any integrated photonic platform and the first assumption-free statistical validation of integrated squeezing performance. Our results establish thin-film lithium niobate as a high-performance, scalable platform for continuous-variable quantum sensing, communications, and computing.
\end{abstract}

\section{Introduction}\label{sec1}
The generation of quantum squeezed light enables the suppression of noise in one field quadrature below the shot-noise limit imposed by Heisenberg’s uncertainty principle~\cite{yuen1976two,walls1983squeezed}. Over the past decade, squeezed light has significantly enhanced the sensitivity of gravitational-wave detectors~\cite{ganapathy_broadband_2023, zhao_frequency-dependent_2020},\hide{virgo_collaboration_frequency-dependent_2023,} highlighting its importance for practical quantum sensing applications~\cite{zhang2021distributed}. Furthermore, replacing coherent vacuum with squeezed states has enabled performance improvements in a wide range of fields, including quantum imaging~\cite{casacio_quantum-enhanced_2021, taylor_biological_2013}, quantum computing \cite{CV_computing,aghaee_rad_scaling_2025,konno_logical_2024} \hide{not comuniaction, ~\cite{zhang2021distributed,guo2020distributed,xia2023entanglement}}, dual-comb spectroscopy~\cite{herman_squeezed_2025,hariri2025} and opto-mechanical sensing \cite{xia2023entanglement,yue_optomechanical} .

Quantum squeezing has been realized using several nonlinear mechanisms, including quadratic $\chi^{(2)}$ nonlinearities~\cite{vahlbruch_detection_2016,kanter_squeezing_2002,eckstein_highly_2011,stefszky_waveguide_2017, mondain_chip-based_2019,Kashiwazaki_wg_2020,10dB_squeezing_measurement_2026, Bright_pulsed_squeezing_2026, amir_sqz,arge_demonstration_2025,our_squeezing_2026}, Kerr $\chi^{(3)}$ nonlinearities~\cite{dutt_-chip_2015,3.5dB_sqz_kerr_MRR,2.5dB_sqz_kerr_MRR,squeezed_microcomb,sqz_microcomb_Yu_Xi,Xanadu_squeezing}, and optomechanical interactions~\cite{Optomechanical_squeezing_2020}. Among these, $\chi^{(2)}$ platforms offer the highest nonlinear light-matter interaction strength and are therefore particularly attractive for generating strong squeezing. State-of-the-art squeezing experiments still predominantly rely on bulk, discrete optical parametric oscillators (OPOs), which have achieved squeezing levels as high as 15 dB via combining $\chi^{(2)}$ nonlinear materials with free-space cavities to provide Purcell enhancement~\cite{vahlbruch_detection_2016, schnabel_squeezed_2017}. 

Photonic integrated circuits (PICs) offer significant advantages in size, weight, and power (SWaP) compared with free-space systems, providing a promising route toward scalable squeezed-light sources. More importantly, tight optical confinement enhances light–matter interaction and enables dispersion engineering, potentially allowing efficient squeezing even without cavity enhancement. \hide{Photonic integrated circuits (PICs) offer significant advantages in size, weight, and power (SWaP) compared with free-space systems, providing a promising route toward scalable squeezed-light sources. Lithium niobate (LN), which possesses one of the largest second-order nonlinear coefficients, has emerged as a leading material for optical parametric amplification (OPA) and squeezed-light generation. } In particular, the thin-film lithium niobate (TFLN) has emerged as a leading material platform for optical parametric amplification (OPA) and squeezed-light generation by leveraging tight modal confinement and large second-order nonlinear coefficients \cite{2600_LN_waveguide,lu_ultralow-threshold_2021,kellner_low_2025}. Although squeezing has recently been demonstrated in TFLN resonators~\cite{amir_sqz,arge_demonstration_2025,our_squeezing_2026,amir_sqz}, achieving simultaneously high escape efficiencies and low propagation losses remains challenging. In addition, maintaining the stringent co-resonance condition at both the pump and squeezed wavelengths requires precise temperature stabilization \cite{our_squeezing_2026, ran_noise_paper,Xinyi_PR_paper}, which has limited the inferred squeezing levels to less than 8 dB \cite{our_squeezing_2026}.

In contrast, non-resonant periodically poled thin-film lithium niobate (PP-TFLN) waveguide-based approaches provide a simpler and monolithic architecture capable of generating broadband squeezing~\cite{nehra_few-cycle_2022,chen_ultra-broadband_2022,stokowski_integrated_2023,shi_squeezed_2025}. Recent experiments have demonstrated 5.6 dB anti-squeezing and $-0.77$ dB squeezing using a 1-cm PPLN waveguide under continuous-wave (CW) pumping~\cite{shi_squeezed_2025}, while inferred pulsed squeezing of 10.5 dB has also been reported using pulses source with a 10-W peak power and a 75-fs pulse duration \cite{nehra_few-cycle_2022}. Despite the use of high peak powers and ultrafast optical pulse sources, pulsed squeezing schemes can still involve practical considerations such as requiring a temporally matched local oscillator (LO) for efficient homodyne detection, while the effective interaction length is often influenced by group-velocity matching. The achievable squeezing benefits from stronger second-order nonlinear interaction sustained over a longer interaction length, with the gain scaling approximately as $\sqrt{gPL}$, where $g$ is the nonlinear coupling coefficient, $P$ is the pump power, and $L$ is the interaction length. However, maintaining strong nonlinear coupling over long waveguides is challenging because thickness variations can disrupt phase matching and reduce conversion efficiency~\cite{Linran_adaptive_poling_2024,wafer-scale_poling}. More importantly, quantum-limited noise variance reduction (namely squeezing) is known to be highly vulnerable to any optical loss, which increases also with interaction length L, posting additional challenges compared with achieving large classical parametric gain. Demonstrating strong squeezing therefore requires excellent device performance together with accurate modeling to confidently extract the true squeezing level~\cite{wu2026BNOT}. In this context, low-loss, long, adaptively poled PPLN nanowaveguides offer a promising route toward achieving continuous-wave squeezing levels beyond 10 dB, a milestone not yet demonstrated on any integrated platform, while approaching the performance of state-of-the-art bulk OPO systems and maintaining broadband operation with reduced system complexity.

In this work, we demonstrate CW quadrature squeezing in a 1.6-cm-long adaptively poled TFLN waveguide with a low propagation loss of 0.1 dB/cm. By mapping the lithium niobate thickness prior to poling, we tailor the local poling period to maintain strong parametric gain across the entire device length, achieving a second-harmonic generation (SHG) efficiency of \SI{2740}{\percent \ W^{-1}}. Our device exhibits the highest reported anti-squeezing for any waveguide or resonator platform, reaching +13.22 $\pm$ 0.44 dB under CW operation at a pump power of 200 mW, along with measured squeezing of -0.66 $\pm$ 0.35 dB. Leveraging a bidirectional SHG method to accurately quantify facet losses, we establish a comprehensive loss and phase-noise model that captures the distributed nature of squeezing generation. Using this framework, we infer a record on-chip squeezing of -18.06 $\pm$ 0.43 dB and anti-squeezing of +20.64 $\pm$ 0.35 dB among all waveguide and resonator-based platforms. At the 95$\%$ confidence level, the on-chip squeezing lies in the range [-18.96, -17.25] dB. To our knowledge, this is the first work to employ a distributed and bidirectional model to infer on-chip squeezing with such high accuracies. Notably, at such high squeezing levels, even sub-percent optical losses and phase noise on the order of tens of milliradians are found to play a non-negligible role, highlighting the critical importance of distributed loss and noise engineering in traveling-wave squeezing systems.

%%%%%%%%%%%%%%%%%%%%%%%%
\begin{figure}[H]
    \centering
    \includegraphics[width=\textwidth]
    {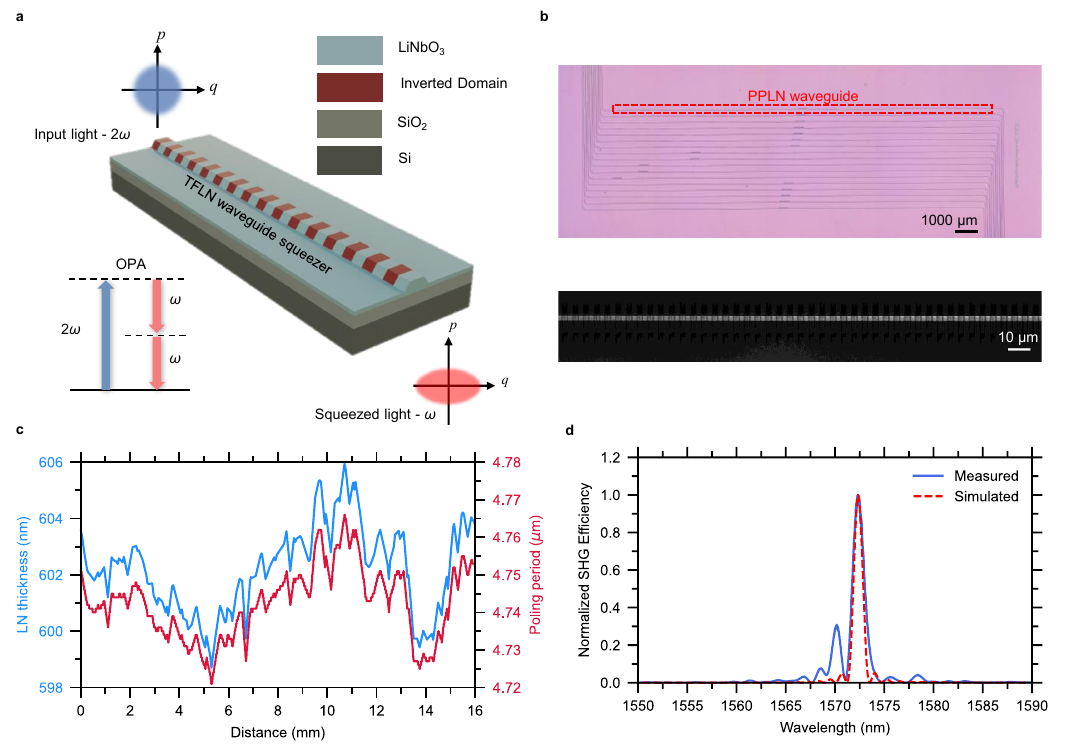}
    \caption{Concept and design of the TFLN PPLN waveguide for on-chip squeezed-vacuum generation.
    \textbf{a}, Schematic of the PPLN nanowaveguide generating squeezed light at frequency $\omega$ (red, output) from a pump at $2\omega$ (blue, input) via a $\chi^{(2)}$ nonlinear interaction.
    \textbf{b}, Optical micrograph of the squeezer chip containing multiple adaptively poled waveguides of length \SI{1.6}{cm} (highlighted by the red dashed box). The lower panel shows a two-photon microscope image revealing the periodically poled domains.
    \textbf{c}, Blue: measured thickness variation along the nanowaveguide obtained via spectroscopic reflectometry. Red: corresponding adjustment of the poling period along the waveguide length to preserve the QPM condition at a fixed wavelength.
    \textbf{d}, Measured SHG spectrum (solid line) of the PPLN waveguide showing a single dominant peak enabled by the adaptive poling design, along with the simulated spectrum (dashed line). Note that both curves are normalized with respect to their own peak efficiency.
    }
    \label{fig: concept}
\end{figure}
%%%%%%%%%%%%%%%%%%%%%%%%

\section{Results}\label{sec2}

We generate squeezed vacuum using a straight PPLN waveguide, as shown in \cref{fig: concept}a. The device is fabricated on a 600-nm-thick MgO-doped X-cut TFLN wafer. \cref{fig: concept}b shows an optical microscope image of the squeezer chip featuring multiple waveguides with \SI{1.6}{cm}-long PPLN sections. The lower panel of \cref{fig: concept}b shows a two-photon microscope image of a PPLN waveguide, revealing uniform domain inversion along the entire length.

Although longer poling lengths can, in principle, yield higher squeezing, nanometer-scale thickness variations in the LN film limit the nonlinear efficiency of TFLN waveguides. To mitigate this, we first map the TFLN wafer thickness and then adjust the local poling period to maintain a constant quasi-phase-matching (QPM) wavelength, as illustrated in \cref{fig: concept}c. The waveguides are fabricated by etching \SI{350}{nm} of LN after periodic poling. After cleaving the chip to expose the facets, a \SI{200}{nm}-thick oxide cladding is deposited to reduce coupling losses. More details on the fabrication procedure can be found in the supplementary material SM Sec. I.

\begin{figure}[H]
    \centering
    \includegraphics[width=0.5\textwidth]
    {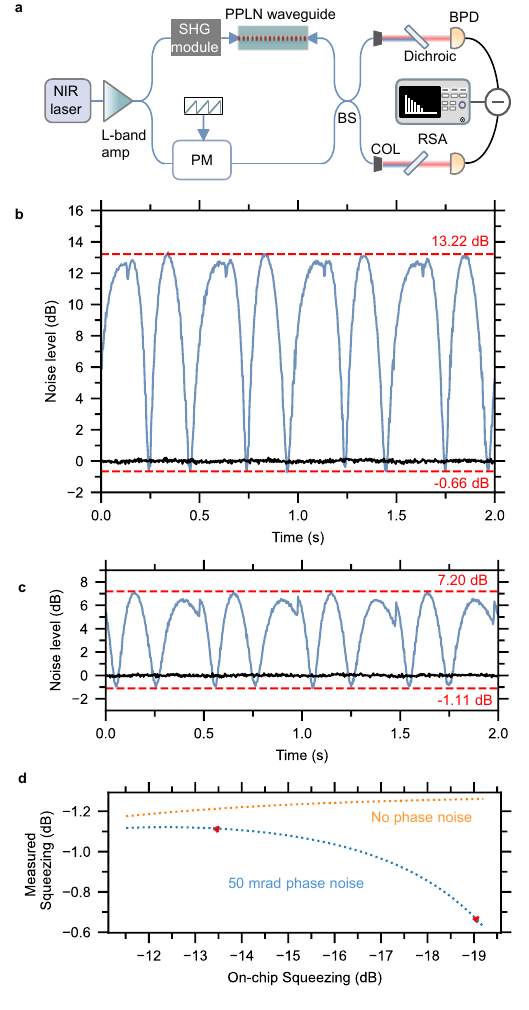}
    \caption{Observation of squeezing from the PPLN nanowaveguide.
    \textbf{a}, Experimental setup for the squeezing measurement. The NIR laser is amplified and split into two paths for squeezing generation and the local oscillator (LO), which are subsequently recombined at the balanced homodyne detector. PM: phase modulator; COL: collimator; BS: beam splitter; BPD: balanced photodetector; RSA: radio-frequency spectrum analyzer.
    \textbf{b}, Normalized quantum noise (blue) measured at \SI{205}{mW} pump power while sweeping the squeezing angle. The black curve indicates the vacuum noise level obtained by blocking the signal path.
    \textbf{c}, Normalized quantum noise (blue) measured at \SI{87}{mW} pump power. 
    \textbf{d}, Measured squeezing vs. on-chip squeezing for \SI{50}{mrad} laser phase noise and no phase noise. Experimental data is represented by red dots. 
    }
    \label{fig: squeezing}
\end{figure}

The PPLN waveguide width is chosen to be \SI{2}{\um} in order to maintain low propagation loss at both telecom and visible wavelengths. We measure the intrinsic quality factors of the PPLN microresonators to be $3.2$ million and $2.0$ million around \SI{1570}{nm} and \SI{785}{nm}, corresponding to propagation losses of $0.12$ and $0.40$ dB/cm, respectively. The effectiveness of the adaptive poling design is verified by the SHG spectrum shown in \cref{fig: concept}d, which exhibits minimal deviation from the ideal sinc-squared response.               
\Cref{fig: squeezing}a shows the optical characterization setup used to measure squeezing. A near-infrared (NIR) laser is amplified and subsequently split into two paths. The first path is directed to a bulk PPLN SHG module to generate the pump light at 2$\omega$ (786.1~nm), while the second path serves as the local oscillator. Squeezed light at $\omega$ (\nir) is generated in the PPLN waveguide through an optical parametric amplification process. For homodyne detection, the squeezed field is interfered with the LO using a fiber beam splitter connected to balanced photodetectors (BPD), achieving an interference visibility of 97 $\%$. The noise variance of the squeezed vacuum is recorded using a radio-frequency spectrum analyzer (RSA), while the LO phase is scanned using a phase modulator (Thorlabs, LN65S-FC) driven by a \SI{20}{Hz} triangular waveform.

\Cref{fig: squeezing}b shows the measured noise of the squeezed light at a sideband frequency of \SI{5}{MHz}. The signal is normalized to the shot-noise level, obtained by blocking the squeezed-light path. We observe $\maxsqz{} \pm \maxsqzerr{}$ of squeezing and $\maxasqz{} \pm \maxasqzerr{}$ of anti-squeezing at an input on-chip pump power of \pump{}. The relation between on-chip squeezing/anti-squeezing and the measured value is given by:
\begin{equation}
\begin{aligned}
G_{\mathrm{\pm},meas}
&=(\eta G_{\mathrm{\pm},onchip}+1-\eta)\cos^{2}(\Delta\theta) \\
&\quad +(\eta G_{\mathrm{\mp},onchip}+1-\eta)\sin^{2}(\Delta\theta),
\end{aligned}
\label{eq:meas_sqz}
\end{equation}
where $G_{-}$ and $G_{+}$ are the squeezing and  anti-squeezing level, respectively, $\eta$ is the total effective detection efficiency and $\Delta\theta$ characterizes the phase noise in the interferometer between the squeezing generator path and the LO path. Consequently, the measured squeezing is primarily limited by detection losses, including the output chip facet loss and losses from subsequent optical components. The relative phase noise in the fiber interferometer further lowers the measured squeezing due to the laser phase noise fluctuation, fiber length fluctuation, as well as any path length difference between the squeezed and LO signals. Phase noise causes the anti-squeezed quadrature to mix into the measured squeezed quadrature, thereby substantially degrading the observed squeezing at high squeezing levels. Consequently, the maximum measured squeezing of -1.11 dB occurs at a lower pump power, as shown in \cref{fig: squeezing}c. \cref{fig: squeezing}d illustrates this effect. By incorporating \SI{50}{mrad} of laser phase noise into \cref{eq:meas_sqz}, we reproduce the reduction in observed squeezing at higher pump powers.

\begin{figure}[H]   % the * makes it span both columns
    \centering
    \includegraphics[width=\textwidth]
    {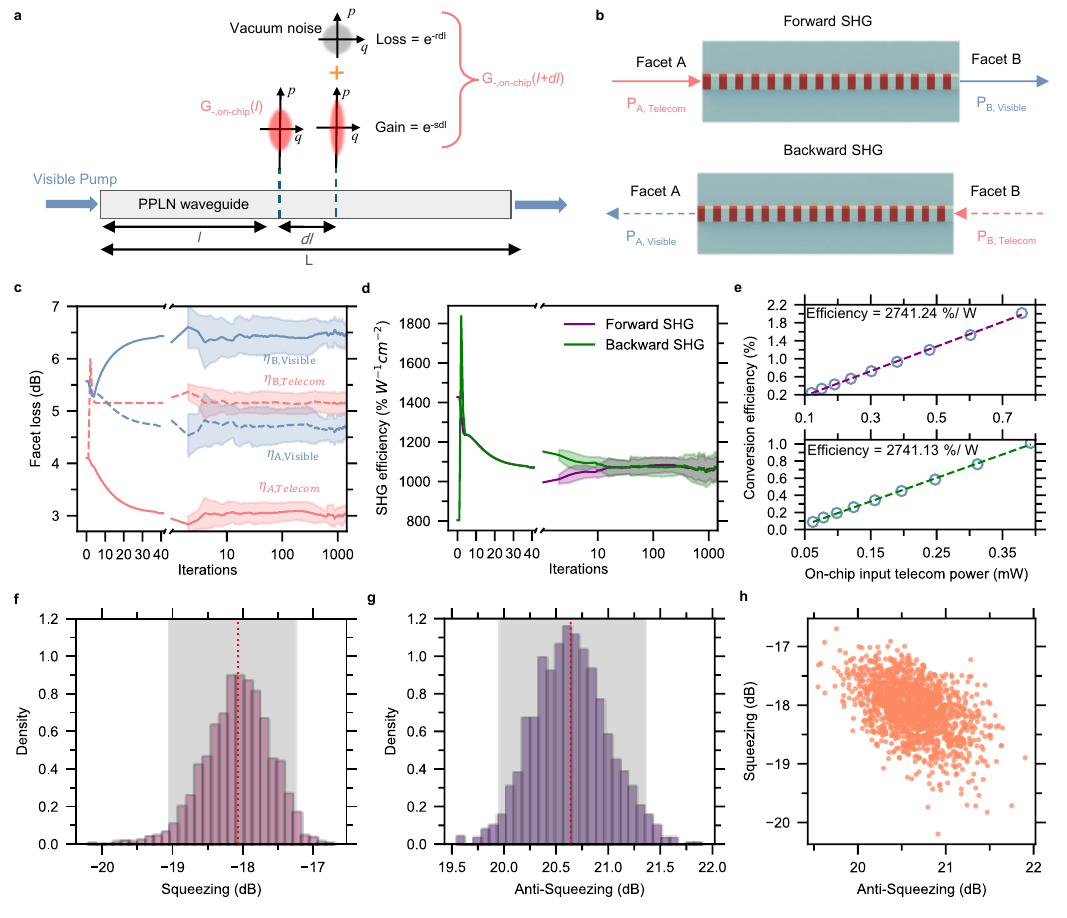}
    \caption{Prediction of on-chip squeezing/anti-squeezing using non-linear equation solver. 
    \textbf{a}, Evolution of continuous-wave squeezing along the PPLN waveguide. As the field propagates, nonlinear interaction generates squeezing while optical loss introduces vacuum noise, progressively mixing with the squeezed state.
    \textbf{b}, Schematic of the bidirectional SHG measurement used to accurately characterize the telecom and visible coupling losses at both facets.
    \textbf{c}, Iteration history of the four unknown facet losses. Starting from an initial assumption of symmetric losses, the parameters are iteratively updated using a nonlinear solver until the residual falls below a predefined tolerance, indicating convergence. The iterations to the right of the x-axis break include Monte Carlo simulations, where the solid line represents the floating-point average and the shaded region denotes the statistical spread.
    \textbf{d}, Evolution of the SHG efficiency in the forward (purple) and backward (green) direction calculated by using the facet loss values in c.
    \textbf{e}, Linearly fitted SHG efficiency for forward propagation (top) and the backward propagation (bottom) using the final facet losses obtained from the non-linear solver.
     Probability density distribution of the inferred squeezing values \textbf{f}, and anti-squeezing \textbf{g}, with the shaded region indicating the 95\% confidence interval and the dotted line denoting the mean.
     \textbf{h}, Joint distribution of inferred squeezing and anti-squeezing from all Monte Carlo trials, illustrating the correlation between the two inferred parameters arising from propagated experimental uncertainties.
    }
    \label{fig: Monte_carlo}
\end{figure}

While optical component loss can be measured directly using sensitive power meters, chip facet loss and interferometer phase noise are fundamentally  challenging to determine since one can not simply measure the optical transmission before and after a certain chip facet and the interferometer involves frequency conversion in the squeezing generation arm,  which can lead to inaccurate estimation of $G_{\pm,onchip}$. Moreover, we note that on chip propagation loss is no longer negligible in a high squeezing regime since it mixes vacuum noise as the squeezed light is generated and propagates along the waveguide (see \cref{fig: Monte_carlo}a). In previous demonstrations, this effect has typically been approximated as a lumped loss combined with off-chip losses, which can lead to significant errors in inferred squeezing values. Here, we adopt a distributed loss-and-squeezing model to accurately capture these dynamics. For a straight PPLN waveguide of length $L$ with SHG efficency $\eta_{SHG}$, the on-chip squeezing/anti-squeezing for incident power $P_{pump,onchip}$  at $2\omega$ is expressed using our model:
\begin{equation}
\begin{aligned}
G_{\mathrm{\mp},onchip} = \frac{{r}\pm se^{-({r}\pm s){L}}}{{r}\pm s}
\end{aligned}
\label{eq:onchip_sqz}
\end{equation}
where $s = 2\sqrt{\eta_{SHG}\eta_{A,v}P_{pump}}$ is the squeezing gain and $r$ is the propagation loss per unit length along the waveguide at $\omega$. The detailed derivation can be found in the the supplementary material SM Sec. V. We extract a waveguide propagation loss of $\proploss \pm 0.02$ dB/cm, extracted from quality factor measurements of PPLN microresonators fabricated under identical process conditions (2.5-3.5 million at near 1570 nm). We also observe similar intrinsic quality factors between PPLN microresonators and non-poled TFLN microresonators with identical geometries, indicating that the poling process does not introduce measurable additional propagation loss.

To avoid uncertainty in estimation of individual facet losses and phase noise, we propose a robust model to evaluate the essential parameters needed to infer on-chip squeezing. Our method relies on three key elements: measuring the chip transmission, deriving theoretical model for on-chip squeezing, and performing bi-directional SHG efficiency measurements.

To track edge coupling losses, we label the input and output facets as A and B, respectively. Coupling losses at telecom (t) and visible (v) wavelengths are denoted $\eta_{A,t}$, $\eta_{B,t}$, $\eta_{A,v}$ and $\eta_{B,v}$. The total detection efficiency for the squeezing measurement in \cref{eq:meas_sqz} is $\eta = \eta_{B,t}\eta_{comp}$, where $\eta_{comp}$ accounts for the losses in optical components after the chip. The total transmission loss of the squeezer chip at $\omega$ (\nir) and $2\omega$ (\visible) is  expressed in terms of the facet loss:
\begin{equation}
\begin{aligned}
\eta_{A,t}\eta_{B,t}
= 10^{-\mathrm{Loss}_t/10}
\end{aligned}
\label{eq:facet_solutiont}
\end{equation}

\begin{equation}
\begin{aligned}
\eta_{A,v}\eta_{B,v}
= 10^{-\mathrm{Loss}_v/10}
\end{aligned}
\label{eq:facet_solutionv}
\end{equation}
where $Loss_{t}$ and $Loss_{v}$ are measured in dB.

%%%%%%%%%%%%%%%%%%%%%%%%
%%%%%%%%%%%%%%%%%%%%%%%%
%%%%%%%%%%%%%%%%%%%%%%%%

To measure $\eta_{SHG}$, we used an optical fiber setup, where a tunable NIR laser output is coupled into the squeezer chip and the converted second-harmonic signal is measured using a visible photo detector (Thorlabs PDA100A2) after WDM filtering. The on-chip SHG efficiency is characterized for both forward and backward propagation, as shown in \cref{fig: Monte_carlo}b. In the forward direction, the fundamental wave ($\omega$) is injected via facet A, and the second-harmonic signal ($2\omega$) is monitored at facet B. In the reverse configuration, the input and collection ports are swapped to assess the symmetry. The SHG efficiency in forward ($\eta_{\mathrm{SHG_{f}}}$) and backward ($\eta_{\mathrm{SHG_{b}}}$) directions is calculated as:
\begin{equation}
\begin{aligned}
\eta_{\mathrm{SHG_{f}}} =\frac{{P_{B,v}}/\eta_{B,v}}
{\big({P_{A,t}}\eta_{A,t}\big)^2}
,
\\
\eta_{\mathrm{SHG_{b}}} =\frac{{P_{A,v}}/\eta_{A,v}}
{\big({P_{B,t}}\eta_{B,t}\big)^2}
,
\end{aligned}
\label{eq:SHG_measurement}
\end{equation}
where the $P_{A,t}$ and $P_{B,v}$ are the incident telecom and output visible power in forward propagation and $P_{B,t}$ and $P_{A,v}$ represent the incident telecom and ouptut visible power in backward propagation. Since the on-chip SHG is theoretically reciprocal, $\eta_{\mathrm{SHG_{f}}}$ should equal $\eta_{\mathrm{SHG_{b}}}$. \Cref{eq:meas_sqz,eq:facet_solutiont,eq:facet_solutionv,eq:onchip_sqz,eq:SHG_measurement} form a set of non-linear equations that we solve to determine the chip facet losses and the phase noise. 
We implement a Levenberg-Marquardt (LM) algorithm for iterative solution, starting with an initial guess of symmetric facet losses. \cref{fig: Monte_carlo}c shows the evolution of the facet losses at each iteration. The corresponding SHG efficiencies in forward and backward directions converge to a single consistent value, as shown in \cref{fig: Monte_carlo}d. Monte Carlo simulations embedded within the solver provide uncertainty estimates for all parameters, yielding an SHG efficiency of $\SHG \pm \SHGerr$.
The converged solutions are provided in \cref{tab:solver_results}. All other parameters including $Loss_{t}$, $Loss_{v}$, $P_{A,t}$, $P_{B,t}$, $P_{A,v}$, and $P_{B,v}$ in these equations are direct power measurements with known error bars (see SM Sec. III). Notably, our model does not assume symmetric facet losses, and instead predicts realistic asymmetric coupling at the input and output facets across both fundamental and second-harmonic wavelength bands. This enables accurate extraction of the on-chip SHG efficiency (in \%/W/cm$^{2}$), which is otherwise highly sensitive to facet-loss assumptions, and provides a reliable means to assess the true poling quality.

\begin{table}[htbp]
\centering
%\captionsetup{width=\textwidth}
\caption{Predicted values by non-linear equation solver at 205 mW pump power}
\label{tab:solver_results}

\begin{tabular}{ccc}
\hline
Parameter & Predicted Mean & \shortstack{95\% \\ confidence interval}  \\
\hline
$\eta_{A,t}$ (dB)  & 3.02& [2.68, 3.38] \\
$\eta_{A,v}$ (dB)  & 4.68&[4.77, 5.56] \\
$\eta_{B,t}$ (dB)  & 5.15& [4.19, 5.25]\\
$\eta_{B,v}$ (dB)  & 6.46& [5.87, 6.95]\\
$\Delta \theta$ (mrad)  & 37 &[2.3 , 55]\\
$\eta_{SHG}$ ($\frac{\%}{W cm^{2}}$)  & 1065.35 &[947.212, 1165.64]\\
$G_{\mathrm{-},onchip}$ (dB)  & -18.06 &[-18.96, -17.25]\\
$G_{\mathrm{+},onchip}$ (dB)  & 20.64 &[19.96, 21.35]\\
\hline
\end{tabular}

\end{table}

The lower measured efficiency compared with our theoretical prediction of 3200 $\% \ W^{-1} cm^{-2}$ (calculated by using analysis provided in \cite{2600_LN_waveguide}) is attributed to incomplete poling depth (supplementary material SM Sec. II) and cladding non-uniformity, which also explains the deviations from the ideal sinc-squared SHG spectrum in \cref{fig: concept}e. \cref{fig: Monte_carlo}e shows the percentage conversion into the second-harmonic for both forward and backward propagation directions as a function of input telecom power. Additionally, the predicted laser phase noise is $0.037 \pm 0.011$ radian which is close to the measured value of 0.062 radian from an independent experiment using a similar interferometer without
\newpage
%%%%%%%%%%%%%%%%%%%%%%%%
%%%%%%%%%%%%%%%%%%%%%%%%
the squeezing generation (supplementary material SM Sec. VII). The squeezing/anti-squeezing used to compute the facet losses in \cref{fig: Monte_carlo}c were obtained at the highest pump power. 

Using this method we infer a maximum on-chip squeezing of $\maxsqzonchip \pm \maxsqzonchiperr$ and on-chip anti-squeezing of $\maxasqzonchip \pm \maxasqzonchiperr$ at 205 mW pump power. The full probability distributions of the inferred squeezing and anti-squeezing obtained from Monte Carlo simulations are shown in \cref{fig: Monte_carlo}(f–h). Detailed solutions for multiple power measurements can be found in the supplementary material SM Sec. VI. \cref{fig: sqzfitting}a shows the measured squeezing and antisqueezing at various pump powers, alongside the theoretical predictions of $G_{\mathrm{\pm},onchip}$ and $G_{\mathrm{\pm},meas}$ using solved facet losses and phase noise. The experimental results show excellent agreement with the model across the full pump-power range.

%%%%%%%%%%%%%%%%%%%%%%%%
%%%%%%%%%%%%%%%%%%%%%%%%
\begin{figure}[H]
\begin{center}

    \includegraphics[width=0.5\textwidth]{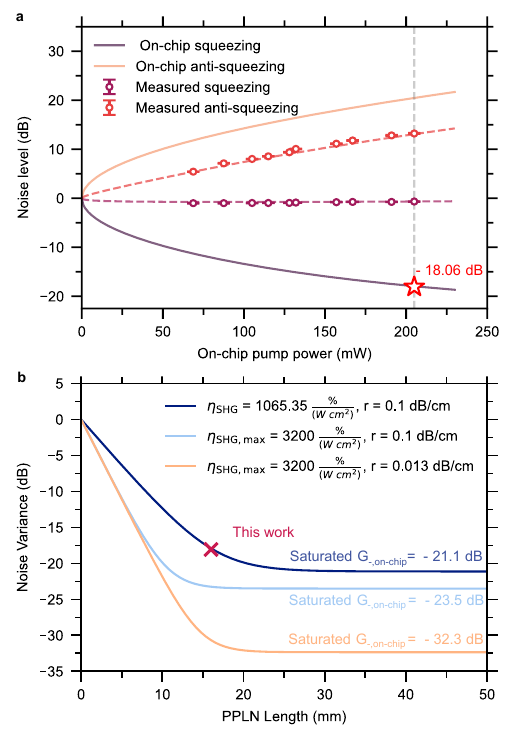}
    \caption{ Estimation of on-chip squeezing using the distributed loss model.
    \textbf{a},  Measured squeezing and anti-squeezing values (circles) at different pump powers are plotted along with the theoretical model (dashed line) incorporating the estimated SHG conversion efficiency, various loss mechanisms, and relative phase noise. From this analysis, we infer a maximum on-chip squeezing of $\maxsqzonchip{} \pm \maxsqzonchiperr{}$ and anti-squeezing of $\maxasqzonchip{} \pm \maxasqzonchiperr{}$ (solid lines).
    \textbf{b}, Predicted squeezing vs. waveguide length using distributed loss model for $\eta_{SHG}$ = 1065.35 $\% \ W^{-1} cm^{-2}$ and 0.1 dB/cm loss (current device), $\eta_{SHG}$ = 3200 $\% \ W^{-1} cm^{-2}$  (theoretical maximum) and 0.1 dB/cm loss, $\&$  $\eta_{SHG}$ = 3200 $\% \ W^{-1} cm^{-2}$ and 1.3 dB/m loss (improved fabrication).  }
    \label{fig: sqzfitting}
\end{center}
\end{figure}
%%%%%%%%%%%%%%%%%%%%%%%%

\section{Conclusion}\label{sec3}
To summarize, we have demonstrated a promising approach for generating on-chip quantum squeezing using TFLN- based PICs. Our PPLN waveguide achieves measured/on-chip continuous wave squeezing and anti-squeezing of -0.66/-18.06 dB and +13.22/+20.64 dB at telecommunication wavelength near 1570 nm using \SI{205}{mW} pump power. Using a non-linear coupled equation solver combined with Monte Carlo simulations, we infer an on-chip squeezing level with a $95\%$ confidence interval of $[-18.96,-17.25]$ dB. Among all integrated platforms, we achieve the highest measured anti-squeezing and the highest inferred on-chip squeezing. The high squeezing gain is enabled by adaptive poling strategy and low propagation loss, which together preserves high non-linear efficiency in a 1.6-cm-long PP-TFLN waveguides. We also demonstrate the longest TFLN waveguide reported to date for use as a non-resonant squeezer. Compared to pulsed pumping schemes, our device achieves higher on-chip squeezing at an order of magnitude lower pump peak power. Furthermore, the OPA gain bandwidth in TFLN can extend over several terahertz, making it a promising platform for broadband and two-mode squeezing protocols.

Looking ahead, extending the current PPLN waveguide length to 2 cm could further improve the squeezing to -19.9 dB, approaching the saturated value of -21.1 dB (\cref{fig: sqzfitting}b). In addition, achieving an SHG efficiency close to the theoretical limit of 3200 $\% \ W^{-1} cm^{-2}$ could enable up to \SI{-23}{dB} of on-chip squeezing. Further improvements in waveguide propagation loss to around \SI{1.3}{dB/m} (corresponding to an intrinsic Q-factor of 30 million \cite{30millionQfactor}) can increase the squeezing to \SI{-32}{dB} (\cref{fig: sqzfitting}b). Further improvements in poling depth through techniques such as side-wall poling \cite{Franken_UV_SPLN_2025} or high-temperature poling \cite{Rajiv_Ram_hightemp_poling} could enhance the achievable squeezing levels. In addition, reducing facet loss using high-efficiency mode size converters \cite{mode_size_converter} would significantly improve the measured squeezing level, which is currently limited by the detection losses. Beyond squeezed light generation, adaptively poled TFLN waveguides can provide parametric amplification for emerging integrated photonic computing architectures \cite{photonic_computing}. Moreover, this non-resonant squeezing approach simplifies the experimental implementation by eliminating the need for precise cavity locking and Pound–Drever–Hall stabilization, which are typically required in resonant systems, thereby making it more suitable for practical deployment. More broadly, this work establishes a powerful framework for extracting both linear and nonlinear device parameters, which have historically been difficult to access using conventional lumped or symmetric-loss models.

\section*{Data availability}
The datasets generated and analyzed in the current study are available
from the corresponding authors on reasonable request.

\section*{Acknowledgements}
This work is supported by the DARPA INSPIRED program (HR001123S0052), the DARPA Young Faculty Award (D23AP00252-02) and the Air Force Office of Scientific Research under award number FA9550-24-1-0349. Two-photon imaging experiments were conducted at the CRL Molecular Imaging Center, RRID: SCR017852, supported by NIH grant S10OD025063.
Device fabrication was performed at the John O’Brien Nanofabrication Laboratory at University of Southern California, Marvell Nanofabrication Laboratory at University of California, Berkeley, University of California Los Angeles Nanolab and Molecular Foundry at Lawrence Berkeley National Laboratory.
M.Y., Y.Y. and T.S.K. are supported by the U.S. Department of Energy, Office of Science, Basic Energy Sciences, Materials Sciences and Engineering Division under Contract No. DE-AC02-05CH11231 within the Quantum Coherent Systems Program KCAS26.
The views, opinions and/or findings expressed are those of the authors and should not be interpreted as representing the official views or policies of the Department of Defense or the U.S. Government.

\section*{Author contributions}
M.Y. conceived the idea. 
T.K. designed the chip with the help of R.K. and C.-H.L.. 
C.-H.L. and T.K. fabricated the devices and developed the fabrication processes with the help of K. K., K. K., C.C., I.C. and Y.Y..
X.R., T.K. and M.C. carried out the experiments and analyzed the data with help from R.K. and J.W. carried out the laser phase noise measurements.

B.-H.W., S.-Y.M., M.J.M., and S.K.V. performed theoretical calculations, supervised by D.E..
Q.Z. developed the theoretical model for inferred squeezing.
T.K., X.R. and M.Y. wrote the manuscript with contributions from all authors. 
M.Y. and Z.C. supervised the project.

\section*{Competing interests}
C.-H.L., Z.C. and M.Y. are involved in developing lithium niobate technologies at Opticore Inc.

%%===========================================================================================%%
\bibliographystyle{unsrtnat}
\bibliography{main}
\end{document}